\documentclass[twoside,slac_one]{revtex4}
\usepackage{graphicx}
\usepackage{fancyhdr}
\usepackage{amsmath} % American Mathematics Society standards
\usepackage{bm}% bold math
\usepackage{amsxtra}
\usepackage{amssymb}
\usepackage{amsthm}
\usepackage{latexsym}
\usepackage{lscape}

\begin{document}

\title{Simulation of Reactors for Antineutrino Experiments Using DRAGON}


\author{L. Winslow}
\affiliation{Department of Physics, Massachusetts Institute of Technology, Boston, MA, USA}

\begin{abstract}
From the discovery of the neutrino to the precision neutrino oscillation measurements in KamLAND, nuclear reactors have proven to be an important source of antineutrinos. As their power and our knowledge of neutrino physics has increased, more sensitive measurements have become possible. The next generation of reactor antineutrino experiments require more detailed simulations of the reactor core. Many of the reactor simulation codes are proprietary which makes detailed studies difficult. Here we present the results of the open source DRAGON code and compare it to other industry standards for reactor modeling.  We use published data from the Takahama reactor to determine the quality of the simulations. The propagation of the uncertainty to the antineutrino flux is also discussed.
\end{abstract}

\maketitle

\thispagestyle{fancy}


%%%%%%%%%%%%%%%%%%%%%%%%%%%%%%%%%%
\section{Introduction}
Reactors are a copious source of antineutrinos, and there is long history of exploiting them for measurements of fundamental neutrino properties. The first reactor antineutrino experiment was in fact the discovery of the neutrino by \cite{ReinesCowan}. More recently, the KamLAND experiment, \cite{KamLAND}, used the reactors of Japan to prove that the effects seen in solar experiments, \cite{SNOIII,SK,Borexino}, is due to a non-zero neutrino mass and the resulting oscillation phenomenon. To complete our understanding of neutrino oscillation, several experiments are using reactors to measure the only unknown mixing angle $\theta_{13}$.  These experiments include Double Chooz, the experiment that motivates this work, \cite{DC2006}, as well as the Daya Bay, \cite{DB2007}, and the RENO, \cite{RENO2010} experiments. Reactors are also good sources for experiments searching for sterile neutrinos, \cite{scraam},  coherent nuclear scattering, \cite{CoherentScatter2004}, and the neutrino magnetic moment, \cite{MUNU2004}.  In addition, there is also a large effort to use antineutrinos to monitor reactors for nonproliferation purposes, \cite{NonProlif2007}. All of these efforts require simulations for their particular reactor cores, and therefore need a code like DRAGON, \cite{DRAGON1994}.

Antineutrinos are produced when fission fragments $\beta$-decay in the reactor core. There are four primary fissile isotopes that produce antineutrinos in the energy range most relevant to experiment. They are $^{235}$U, $^{238}$U, $^{239}$Pu, and $^{241}$Pu, and they produce neutrinos with energies $\lesssim$~8.5~MeV. The flux of antineutrinos that arrives at a detector can be expressed as
\begin{equation}\label{theEquation}
\frac{d^2 n_{\bar{\nu}}}{dE_{\bar{\nu}} dt} = \sum_i^{\text{isotopes}} f_i(t) S_i(E_{\bar{\nu}})
\end{equation}
where $f_i(t)$ is the fission rate of one of the four primary fissile isotopes and $S_i(E_{\bar{\nu}})$ is the spectrum of neutrinos emitted per fission of that isotope. These spectra are extracted from $\beta$-spectrum measurements at the ILL research reactor by \cite{Spec1981,Spec1985,Spec1982,Spec1989}.  It is the re-analysis of these spectra by \cite{Mueller2011} and \cite{HuberReactor} that has led to the ``Reactor Anomaly", \cite{Anomoly2011}.  The DRAGON simulation provides the fission rates, $f_i(t)$, based on a detailed description of the reactor design and operating conditions during a given time period.

To understand the accuracy of the DRAGON predictions, it is necessary to compare the simulation results to data from nuclear reactors. In \cite{Takahama2001}, the Japanese Atomic Energy Research Institute has published the mass inventories of the key isotopes from spent fuel rods used in the Takahama-3 reactor core.  This data has been used to benchmark many widely-used simulation codes. We can use this data to compare DRAGON's results to those from these other codes, and to understand the systematic uncertainties in these results. This work is a subset of the effort presented by \cite{Jones2011}.

%%%%%%%%%%%%%%%%%%%%%%%%%%%%%%%%%%
\section{Reactor Basics}
There are two main types of nuclear reactors: boiling water reactors (BWR) and pressurized water reactors (PWR). Most modern reactors, including Takahama-3, are PWRs. The cores of such reactors are formed by fuel assemblies. Assemblies in turn are made up of fuel rods arranged in a grid. In this grid, some locations are used for instrumentation and some for shaping the neutron flux with special gadolinium rods. A fuel rod is typically 4~m long, and fashioned out of tiny cylindrical fuel pellets measuring 1~cm in diameter and 1~cm in height. The fuel pellet is composed of UO$_2$ enriched to a few percent in $^{235}$U by weight. The structure of the rod is maintained by Zircaloy cladding. Zircaloy is a zirconium alloy chosen for its high melting point. 

The assembly is the elementary unit of fuel in the core.  It arrives at the power plant assembled with fresh fuel, ready to be installed. Reactors refuel approximately once a year. The time between refueling is called a fuel cycle, and during each fuel cycle approximately a third of the fuel assemblies will be fresh. During each refueling, the assemblies are arranged very precisely to create an approximately uniform neutron flux across the core while extracting the most power out of the fuel.

%%%%%%%%%%%%%%%%%%%%%%%%%%%%%%%%%%
\section{DRAGON}
Since the fuel assembly is the building block of the reactor core, it is logical that DRAGON simulates individual assemblies.  To model the whole reactor core with DRAGON, one needs to assume that the spill-in of neutrons from neighboring assemblies matches the loss of neutrons from spill-out. The results presented here indicate that this is good assumption.  For those interested in modeling the full core more accurately, DRAGON results can be interfaced with the full core simulation DONJON, \cite{DONJON2004}. The DRAGON code is open source\footnote{http://www.polymtl.ca/nucleaire/DRAGON/en/index.php} making it attractive for use in antineutrino studies. In fact, small modifications are made in DRAGON for this work to allow the extraction of the fission rates for the antineutrino flux calculation.   

Codes that simulate reactors must solve the neutron transport equations. The deterministic approach uses simplifications to allow the direct solution of these equations. In contrast, Monte Carlo based codes use the random generation of a large neutron sample to solve the equations statistically. DRAGON is a deterministic 2D lattice code. This means that it simulates the assembly by solving the neutron transport equation explicitly on a 2D lattice of fuel cells roughly corresponding to the fuel rods. For a given time step in the evolution of an assembly, the neutron transport equation is solved, and the solution is used to evolve the fuel composition according to the Bateman equations. The deterministic technique is fast compared to Monte Carlo based approaches. Symmetries in the assembly geometry can be exploited to further simplify and accelerate these calculations. 

%%%%%%%%%%%%%%%%%%%%%%%%%%%%%%%%%%
\section{The Takahama Benchmark}
Takahama-3 is PWR reactor located in Japan. Three fuel rods in two fresh assemblies were part of the study outlined in \cite{Takahama2001}. We focus on fuel rod SF97 because it has the largest exposure or burnup, and any cumulative systematic effects will be maximized. The assembly containing SF97 was present in three consecutive fuel cycles of 385, 402, and 406 days with 88 days and 62 days of cool-down time between cycles. At the end of the exposure, six sample disks, 0.5~mm in width, were removed at the positions indicated in Table~\ref{tab:zaxismod}. The samples were dissolved and a chemical separation was performed. Mass spectroscopy was used to determine the mass inventories of these samples. For the four isotopes of greatest interest to us, $^{235}$U, $^{238}$U, $^{239}$Pu, and $^{241}$Pu, the uncertainty is $<$0.1\% for uranium isotopes and $< $0.3\% for plutonium isotopes, ~\cite{Takahama2001}.

\begin{table}
\caption{\label{tab:zaxismod}The position of samples taken from rod SF97,  with the corresponding moderator temperatures, and burnup values for that sample. Measurements are in
  mm from the top of the rod.  The bottom of the rod is at 3863 mm.}
\begin{center}
\begin{small}
\begin{tabular}{| c | c | c | c |}
\hline
 \textbf{Sample}  &   \textbf{Position} &  \textbf{Moderator Temp.} &  \textbf{Burnup} \\
 & [mm] & [K] & [GW-days/ton] \\
\hline
1 & 163 & 593.1 & 17.69 \\
\hline
2 & 350 & 592.8 & 30.73 \\
\hline
3 & 627 & 591.5 & 42.16\\
\hline
4 & 1839 & 575.8 & 47.03 \\
\hline
5 & 2926 & 559.1 & 47.25\\
\hline
6 & 3556 & 554.2 & 40.79 \\
\hline
\end{tabular}
\end{small}
\end{center}
\end{table}

Detailed information on the geometry of the assembly and the basic operation of the reactor is provided. This information encompasses most of the inputs needed to simulate the assembly containing SF97. These inputs are summarized in Table~\ref{tab:inputs}. However, there is insufficient information to simulate the full core, so by definition all Takahama simulations are assembly simulations. The fuel density used in simulations is an effective density to account for the variations in fuel pellet packing.  This effective density must be less than 10.96~g/cm$^{3}$, the theoretical density of UO$_{2}$, but no more details are provided by \cite{Takahama2001}. For this reason, we use the value 10.07~g/cm$^3$ suggested by \cite{Roque2004}.  Another popular assumption is 95\% of the theoretical density of UO$_2$. 

\begin{figure*}
\centering
\includegraphics[width=80mm]{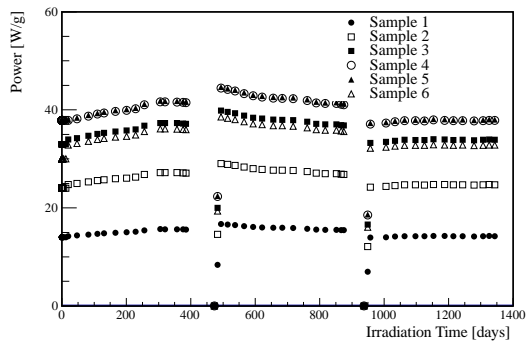}
\caption{The burnup of the six SF97 samples as determined using the $^{148}$Nd technique.} \label{power}
\end{figure*}

\begin{table}
\caption{\label{tab:inputs} Key parameters for the DRAGON simulation.}
\begin{center}
\begin{small}
\begin{tabular}{| c | c |}
\hline
Parameter & Value \\ 
\hline
Moderator Density & 0.72 g/cm$^3$ \\
 \hline
Moderator Temperature & 600.0 K \\ 
 \hline
Cladding Temperature &	600.0 K \\ 
 \hline
Fuel Temperature &	900.0 K \\  
\hline
Fuel Density &	10.07 g/cm$^3$ \\  
\hline
Fuel Cell Mesh & 1.265 cm \\  
\hline
Fuel Rod Radius & 0.4025 cm \\  
\hline
Fuel Cladding Radius & 0.475 cm \\ 
 \hline
Guide Tube Inner Radius & 0.573 cm \\  
\hline
Guide Tube Outer Radius & 0.613 cm \\ 
\hline
\end{tabular}
\end{small}
\end{center}
\end{table}

The last important input to the simulation is the time dependent power of the reactor core. The type of power information provided by the Takahama benchmark is particularly unique. Simulations usually rely on the individual assembly power densities with the full core thermal power. The destructive analysis allows the use of the $^{148}$Nd concentration to produce a detailed power history along the rod, as displayed in Fig.\ref{power}. The  $^{148}$Nd technique has a 3\% uncertainty compared with uncertainties of $<$2\% for the full core thermal power, \cite{Zelimir2009}. The impact of this and other key uncertainties is discussed in the following section.

\section{Results for Rod SF97}
DRAGON uses the description of the fuel assembly from Table~\ref{tab:inputs} to simulate the Takahama assembly. This simulation is performed in time steps with different power and duration as outlined in Fig.~\ref{power}.  To simplify the model, we assume a constant temperature along the rod of 600~K instead of the detailed temperature gradient shown in Table~\ref{tab:zaxismod}.  We also assume a constant non-burnable concentration of boron in the moderator of 630~ppm.  To increase the speed of the calculation, two gadolinium rods are added to the assembly bringing the total to 16. This creates an eight-fold symmetry, and allows the simulation of a 1/8 segment of the assembly.  A nuclear cross-section library must be provided for the calculation.  Our nominal simulation is performed using ENDF/B-VI, \cite{ENDFVI}, and we compare these results to those obtained with JENDL 3.2, \cite{JENDL}. 

\begin{figure*}
\centering
\includegraphics[width=135mm]{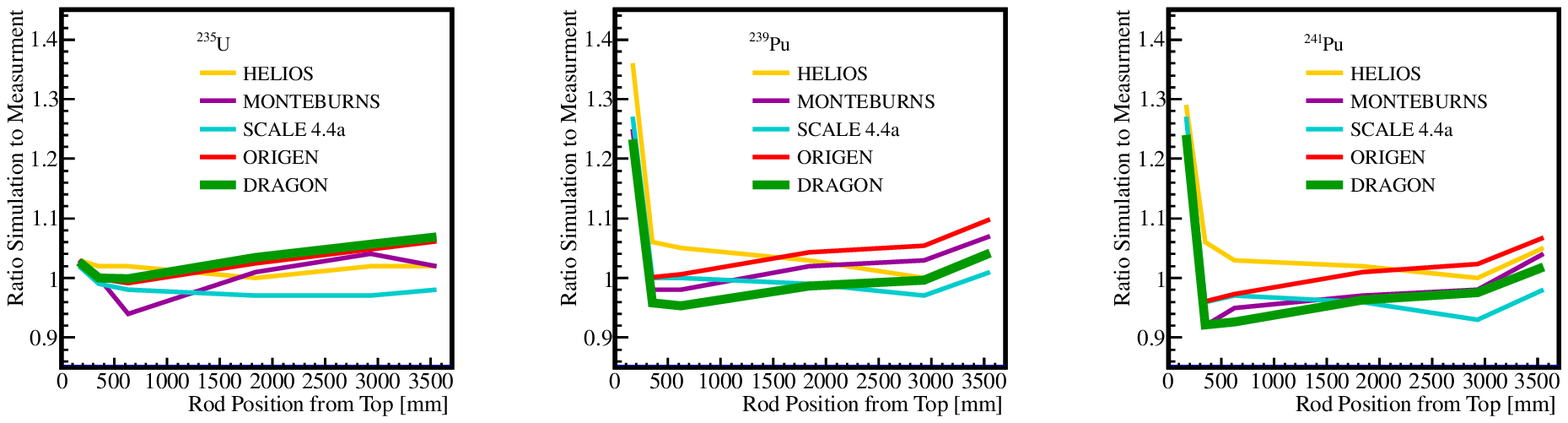}
\includegraphics[width=135mm]{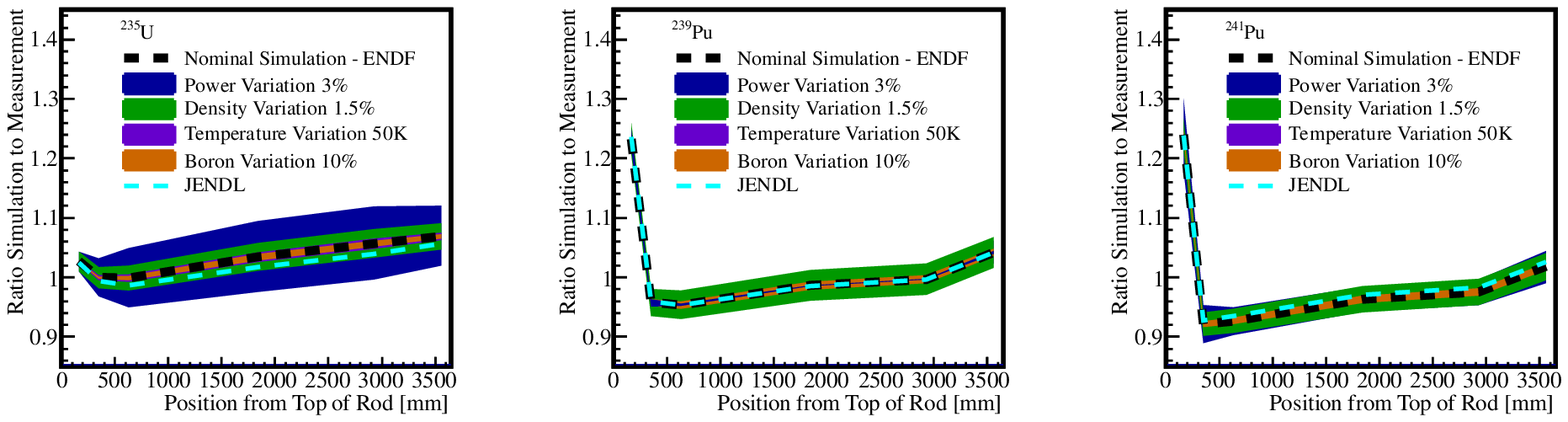}
\caption{The comparison of DRAGON simulation results to the measured mass inventories. Top:  the DRAGON results are compared to other standard codes for reactor core simulation. Bottom: the critical  inputs to the simulation are varied within their uncertainties. } \label{results}
\end{figure*}

The results of this DRAGON simulation are compared to the results from the destructive assay of the fuel rod. These are presented in Fig.~ \ref{results} as a function of the sample's position along the rod. The codes used for comparison range from ORIGEN 2.1, a simple fuel depletion model, ~\cite{ORIGEN},  SCALE4.4a, a code that uses an effective 1D  geometry to model the neutron transport, ~\cite{ScaleHelios1998}, to the full Monte Carlo treatment employed by MONTEBURNS, ~\cite{Monteburns2009}. The code most similar to DRAGON is HELIOS, ~\cite{ScaleHelios1998}, another 2D deterministic lattice code. 

We only present the results for $^{235}$U, $^{239}$Pu, and $^{241}$Pu in Fig.~\ref{results} because the depletion $^{238}$U  is on the same order as the uncertainty in the destructive assay, and therefore a useful comparison is not possible. For sample SF97-1, all codes show substantial deviations in the inventories of the plutonium isotopes. This sample is near the top of the rod. Therefore, the modeling of neutron leakage is difficult leading to these large deviations. Neglecting sample SF97-1, we calculate the average deviation along the rod. For $^{235}$U, DRAGON's deviation is 3.2\%.  For  $^{239}$Pu, and $^{241}$Pu, the deviation is -1.3\% and -3.9\% respectively. For comparison the range for the other codes is -2.2\% to  4.5\% for $^{235}$U, -0.6\% to 6.5\% for $^{239}$Pu, and -4.0\% to 3.4\% for $^{241}$Pu. The DRAGON results have comparable accuracy to those from the other standard codes.

Most work benchmarking reactor simulation codes stops here, but this neglects the uncertainty in these simulated results.  To understand the effect of the input uncertainties on the simulation, we vary key input parameters within their uncertainties. We confirm that these key parameters are the power, moderator temperature, boron concentration and fuel density, as originally outlined in \cite{Zelimir2009}.
We vary the power by 3\%, the uncertainty in the $^{148}$Nd method. We vary the moderator temperature by $\pm$50~K, the variation along the length of the rod, and the boron concentration by 10\%. The fuel density is varied by 1.5\%, but we note that a deviation as large as 3\% is needed to cover the range of values suggested for this effective density.

The largest effect for $^{235}$U is the power variation, a 6\% effect. Since $^{239}$Pu and $^{241}$Pu are the result of fast neutron reactions on $^{238}$U, these isotopes are sensitive to the total amount of fuel. Therefore, they are sensitive to the fuel density, a 2.5\% effect.  The results of these variations are summarized in the bottom of Fig.\ref{results}. If the larger 3\% fuel density uncertainty is assumed, then the DRAGON results are consistent with both the measured mass inventories and the results of the other codes. We also find that the choice of cross-section library induces a 1.2\% change in the $^{235}$U prediction, a 0.8\% change for  $^{241}$Pu and no change for $^{239}$Pu.

The Takahama benchmark focuses on the measured mass inventories. However, the quantities of interest for experiment are the fission rates, $f_i(t)$ in Eq.~\ref{theEquation}, used to calculate the antineutrino flux.  The mass inventories are proportional to the integrated number of fissions and the instantaneous fission rates.  Because the systematic uncertainties on the Takahama inputs are large and the provided power is in a form not available for most reactor cores,  calculating systematic uncertainties on fission rates directly from the Takahama results is not possible. However, it is possible that an experiment could use these results to constrain the uncertainties in the flux calculation in its analysis. Regardless, this work has outlined a procedure for evaluating the major uncertainties and their effect on the fission rates.  This work is applicable to all experiments that would use reactors as their source of antineutrinos.
 
%%%%%%%%%%%%%%%%%%%%%%%%%%%%%%%%%%
\section{Conclusion}
The comparison of measured mass inventories from the Takahama-3 reactor to the DRAGON simulation demonstrates that the results obtained with DRAGON are accurate and of equal quality to other widely used codes for reactor modeling.  We outline the procedure for evaluating the input uncertainties and their effect on the fission rates that are critical to the antineutrino flux calculation. There are many interesting measurements to be done with reactor antineutrinos, and DRAGON is an excellent tool that should be used to aid these measurements.

\begin{acknowledgments}
The author thanks the NSF for their generous support. This work is done in collaboration with the greater Double Chooz Reactor Group, and the author thanks them for their valuable input. The author especially thanks Christopher Jones for the DRAGON simulations presented here.
\end{acknowledgments}

\bigskip % extra skip inserted
\bibliography{dpf2011_takahama}
%\begin{thebibliography}{9}   % Use for  1-9  references
%\begin{thebibliography}{99} % Use for 10-99 references
%\bibitem{charm07}   http://www.lepp.cornell.edu/charm07/
%\bibitem{templates-ref} http://www.slac.stanford.edu/econf/editors/eprint-template/instructions.html
%\end{thebibliography}

\end{document}